\documentclass{PoS}
\usepackage{amssymb}   
\usepackage{amsmath}   
\usepackage{epsfig}
\usepackage{rotating}

\input pix.sty

\def\undertilde#1{\mathop{\vtop{\ialign{##\cr$\textstyle{#1}$\cr%
\noalign{\kern1pt\nointerlineskip}\hfil$\mathchar"0365$\hfil\cr}}}}
\def\wideundertilde#1{\mathop{\vtop{\ialign{##\cr$\textstyle{#1}$\cr%
\noalign{\kern1pt\nointerlineskip}\hfil$\mathchar"0367$\hfil\cr}}}}

\renewcommand{\eq}{eq.~}

\renewcommand{\se}{sec.~}
\renewcommand{\ses}{secs.~}
\renewcommand{\fig}{fig.~}

\newcommand{\tinymsbar}{{\overline{\mbox{\tiny\rm{MS}}}}}
\newcommand{\Lambdamsbar}{{\Lambda_\tinymsbar}}

\newcommand{\Tc}{T_{\rm c}}

\def\lsi{\raise0.3ex\hbox{$<$\kern-0.75em\raise-1.1ex\hbox{$\sim$}}}
\def\gsi{\raise0.3ex\hbox{$>$\kern-0.75em\raise-1.1ex\hbox{$\sim$}}}

\newcommand{\gsim}{\mathop{\gsi}}

\newcommand{\rmii}[1]{{\mbox{\tiny\rm{#1}}}}

\newcommand{\re}{\mathop{\mbox{Re}}}

\newcommand{\Tint}[1]{{\hbox{$\sum$}\!\!\!\!\!\!\!\int\,}_{\!\!\!\!\raise-0.9ex\hbox{$\scriptstyle{#1}$}}}
\newcommand{\Tinti}[1]{{{\Sigma}\!\!\!\!\raise0.3ex\hbox{$\int$}_\rmii{${#1}$}}}

\newcommand{\bi}{\begin{itemize}}
\newcommand{\ei}{\end{itemize}}
\newcommand{\hide}[1]{ }

\title{Towards the continuum limit in transport coefficient computations}

\ShortTitle{Towards the continuum limit in transport coefficient computations}

\author{A.\ Francis$^a$, 
        O.\ Kaczmarek$^b$, 
        M.\ Laine$^c$, 
        M.\ M\"uller${\,}^b$, 
        \speaker{T.\ Neuhaus}${\,}^d$, 
        H.\ Ohno$^e$\\
        \llap{$^a$}
        Institute for Nuclear Physics, 
        JGU Mainz, D-55099 Mainz, Germany\\
        \llap{$^b$}
        Faculty of Physics, University of Bielefeld, 
        D-33501 Bielefeld, Germany\\
        \llap{$^c$}
        Institute for Theoretical Physics, AEC, University of Bern, 
        CH-3012 Bern, Switzerland\\
        \llap{$^d$}
        Institute for Advanced Simulation,  
        FZ J\"ulich, D-52425 J\"ulich, Germany\\
        \llap{$^e$}
        Physics Department, Brookhaven National Laboratory, 
        Upton, NY 11973, USA\\
        E-mail: 
        \email{francis@kph.uni-mainz.de}, 
        \email{okacz@physik.uni-bielefeld.de}, 
        \email{laine@itp.unibe.ch},
        \email{mmueller@physik.uni-bielefeld.de},
        \email{t.neuhaus@fz-juelich.de},  
        \email{hono@quark.phy.bnl.gov}
        }

\abstract{
The analytic continuation needed for the extraction of 
transport coefficients necessitates in principle a continuous
function of the Euclidean time variable. We report on progress
towards achieving the continuum limit for 2-point correlator measurements
in thermal SU(3) gauge theory, with specific attention paid to scale setting. 
In particular, we improve upon the determination of the critical lattice
coupling and the critical temperature of pure SU(3) gauge theory, estimating
$r_0 \Tc \simeq 0.7470(7)$ after a continuum extrapolation. As an application 
the determination of the heavy quark momentum diffusion coefficient from 
a correlator of colour-electric fields attached to 
a Polyakov loop is discussed.
}

\FullConference{31st International Symposium on Lattice Field Theory 
                LATTICE 2013\\
		July 29 - August 3, 2013\\
		Mainz, Germany}

\begin{document}

%
\section{Motivation}

Among quantities playing a central role in the theoretical
interpretation of heavy ion collision experiments at RHIC
and LHC
are so-called transport coefficients: shear and bulk viscosities as well 
as heavy and light quark diffusion coefficients. Because of strong 
interactions, these quantities need to be determined by 
numerical lattice Monte Carlo simulations. 
This represents a challenging problem, 
given that numerical simulations
are carried out in Euclidean signature, whereas transport coefficients
are Minkowskian quantities, necessitating an analytic 
continuation~\cite{harvey_rev}. 
Nevertheless, the problem is solvable 
in principle~\cite{cuniberti}, provided that lattice simulations
reach a continuum limit and that short-distance singularities can be
subtracted~\cite{analytic}. 
The purpose of this investigation is to probe the practical 
feasibility of these steps, by approaching the 
continuum limit for a particular 2-point correlator 
in pure SU(3) gauge theory, related to 
heavy-quark diffusion
(cf.\ \eq\nr{GE_final}). 
An important ingredient in reaching the continuum limit is 
scale setting, so we start by discussing 
issues related to this topic in \ses\ref{se:betac}, \ref{se:Tc}. 

%
\section{Improved determination of the SU(3) phase diagram} %
\la{se:betac}

When SU(3) lattice gauge theory is discretized according to Wilson's
classic prescription, and an infinite spatial-volume limit is taken, the theory
contains two parameters: the number of lattice sites in the temporal 
direction, denoted by $N_\tau$, and the lattice coupling, denoted by 
$\beta$. The phase diagram in the plane ($N_\tau,\beta$) contains
a line of first order transitions; for a given $N_\tau$, 
the critical point is denoted by $\beta_c$. 
The continuum limit corresponds to $N_\tau\to \infty$, $\beta_c\to \infty$.

Because of computational limitations, lattice simulations 
were historically carried out at small values of $N_\tau$. 
In fact, previous to our study, $\beta_c$ had been 
reliably determined only for $N_\tau = 4 - 12$~\cite{bb,ltw}; 
for instance, already at $N_\tau = 16$ the best results available came from
a lattice $ 16 \times 24^3$ which does not 
correspond to the required infinite-volume limit. 
In addition to simulations, 
semi-analytic frameworks have recently been developed
for estimating $\beta_c$~\cite{pos12,pos13}, however these may contain 
uncontrolled uncertainties.

\begin{figure}[t]
    \vspace*{+4.5cm}
  \begin{minipage}[t]{7.0cm}
    \vspace*{-5cm}
    \epsfig{file=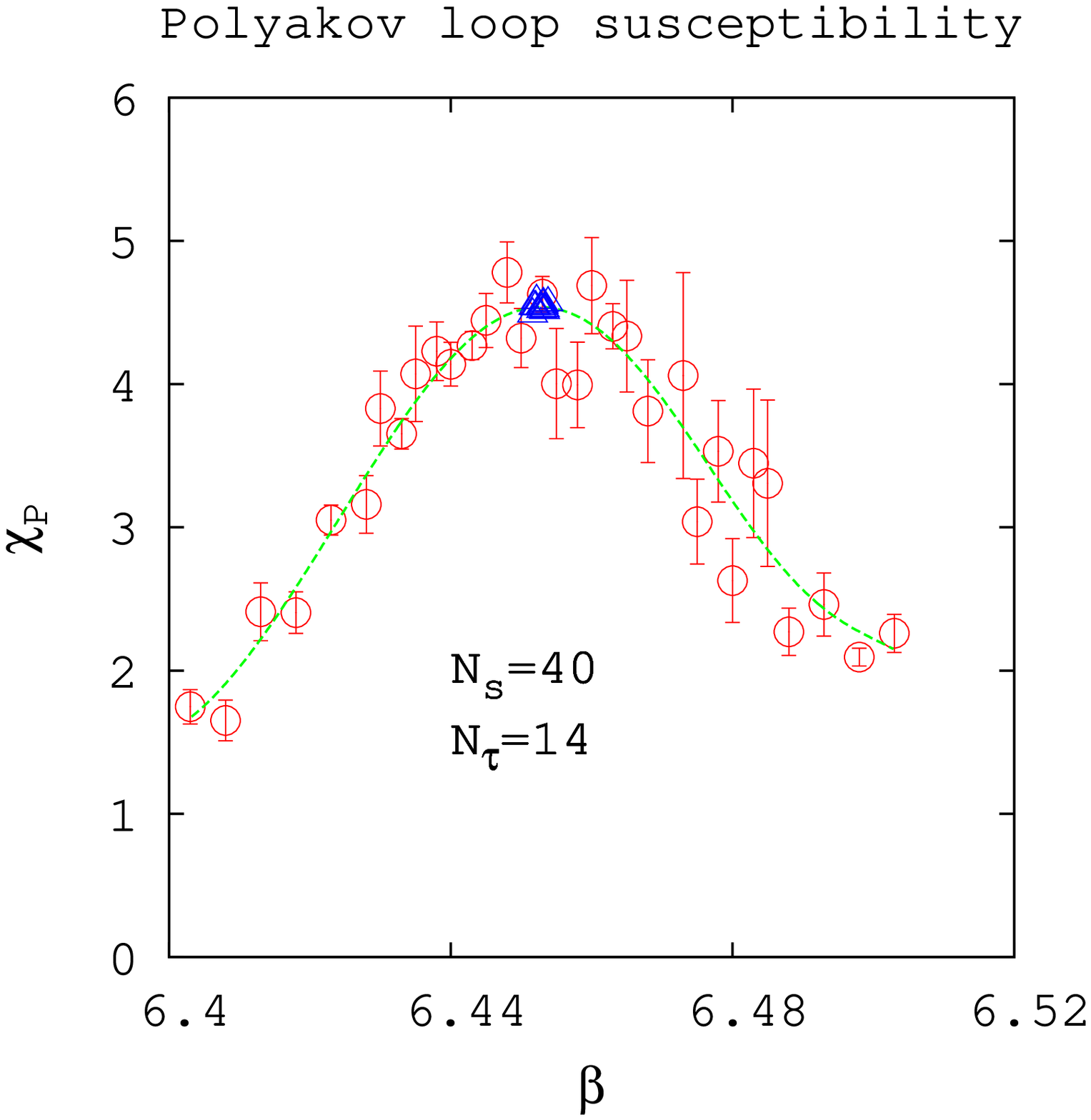,angle=270,width=7.0cm}
  \end{minipage}
  \begin{minipage}[t]{7.5cm}
    \vspace*{-5cm}
    \epsfig{file=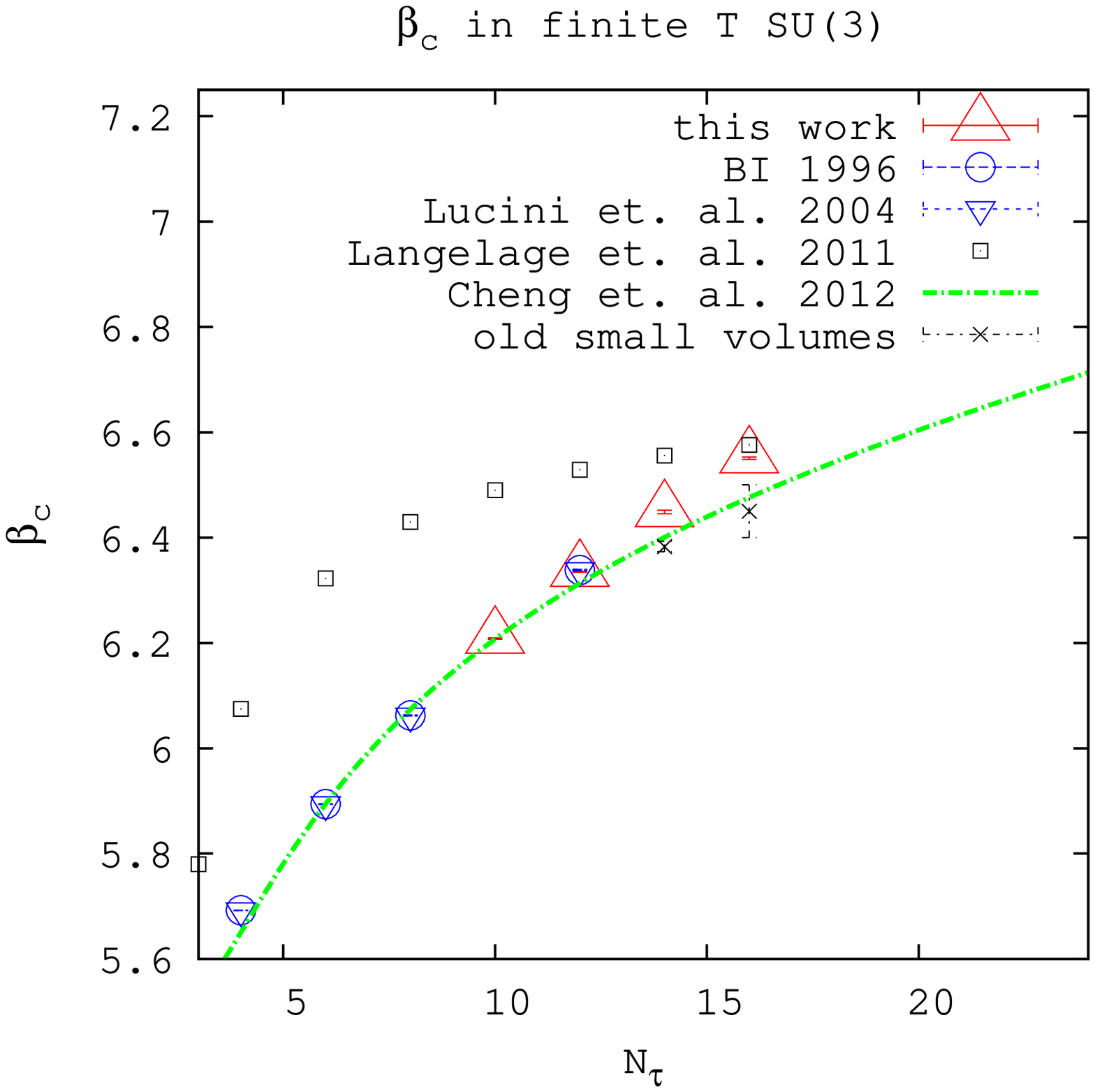,angle=270,width=7.5cm}
  \end{minipage}
\caption[a]{
 Left: 
 Polyakov loop susceptibility $\chi^{ }_\rmii{P} \equiv V_{\rm spatial}
 (\langle P^2\rangle-\langle P\rangle^2)$ for a $14 \times 40^3$ box
 in pure SU(3) lattice gauge theory with the Wilson action.
 Right: 
 Published data for $\beta_c$, from Bielefeld (BI)~\cite{bb} and 
 Lucini et al~\cite{ltw},
 compared with our new data points at $N_\tau=10,12,14,16$. 
 Data from Langelage et al~\cite{pos12} and
 an interpolation to the results of 
 Cheng and Tomboulis~\cite{pos13} represent semi-analytic studies.
 }
\label{fig:betac}
\end{figure}

We have carried out new simulations at $N_\tau = 10, 12, 14, 16$, 
in each case with at least two spatial lattice
sizes, denoted by $N_s$, in the range $N_s \gsim 3 N_\tau$. The critical point 
$\beta_c$ is determined from the peak position of the susceptibility related 
to the Polyakov loop. 
A unique $\Delta \beta=h \times (N_\tau /N_s)^3$ with $h \approx 0.07$
is employed for infinite spatial-volume extrapolations.
For illustration we display in 
\fig\ref{fig:betac}(left) our Polyakov loop 
susceptibility data on a $14 \times 40^3$ lattice 
(at ${N_s / N_\tau} \approx 2.9$) and their 
corresponding Ferrenberg-Swendsen reweighting 
(denoted by the curve in the figure).
Using the peak position of this data (the triangle)
and a similar data set for a $14 \times 56^3$ lattice 
(at ${N_s / N_\tau} =4$) a finite-size extrapolation 
gives $\beta_c(N_\tau=14)=6.4488(59)$.

%
%

In \fig\ref{fig:betac}(right) we display our results for the critical
coupling for $N_\tau=10$, 12, 14 and 16. Also shown are old results 
from refs.~\cite{bb,ltw} 
and the results of semi-analytic computations from  
refs.~\cite{pos12,pos13}. 
Our results for $N_\tau=10$ and 12 agree with the old ones within errors 
whereas for larger $N_\tau$, $\beta_c$
has now been determined relatively reliably
for the first time. 
Even though our current estimate at 
$N_\tau=16$ is preliminary ($\beta_c(N_\tau=16)\simeq 6.5509(39)$), 
it can be seen that the semi-analytic calculations miss some of the 
structure in the data.

%
\section{Conversion of results to physical units} %
\la{se:Tc}

When transport coefficients 
are measured, then simulations are carried out a temperature 
above the critical temperature (denoted by $\Tc$); 
for a fixed $N_\tau$, this corresponds
to $\beta > \beta_c$. For any given $\beta$, it is 
possible to carry out a corresponding zero-temperature simulation, 
on a lattice $N_s^4$, in order to measure some physical quantity. 
Thereby the temperature can be expressed in terms of 
a chosen reference scale; from the results of \se\ref{se:betac}, in turn, 
the reference scale can be determined in terms of $\Tc$.
Expressing everything in units of $T_{\rm c}$ may 
increase the value of the results, given that $T/T_{\rm c}$ 
is a quantity which allows e.g.\ to compare many related theories.

Various reference scales have been used in the past. 
The most traditional one is the root of the string tension, 
denoted by $\sqrt{\sigma}$. The problem is that it is determined
from a fit to the large-distance asymptotics of 
a static potential (or its derivative, the static force), but
such fits are delicate, because the correct 
ansatz needs to be known and also because measurements at large distances
are subtle. Quite concretely, some of the numbers cited by various groups, 
$\Tc/\sqrt{\sigma} = 0.630(5)$~\cite{bb} versus
$\Tc/\sqrt{\sigma} = 0.646(3)$~\cite{ltw}, differ by a value much
larger than the statistical error. 

Another possible scale is the so-called Sommer scale, denoted
by  $r_0$~\cite{rs}. In this case the scale is determined from 
the static force at 
intermediate distances, which removes the need for fitting. 
The price to pay 
is that the discreteness of the lattice hampers
the measurement, and an interpolation together with tree-level
improvement is probably needed for stable results~\cite{rs}. 
A value for $r_0 \Tc$ has been obtained 
in ref.~\cite{sn}: $r_0 \Tc = 0.7498(50)$.

Recently a possible new scale was introduced,  denoted by 
$\sqrt{t_0}$~\cite{ml}. For any given $\beta$, the lattice configuration
is ``cooled'' through a classical Wilson flow until a certain observable
reaches a prescribed value, at time $t_0$. There is no fitting
involved, and no interpolation; $t_0$ therefore probably suffers from less
systematic uncertainties than the other scale choices.

%
\begin{figure}[t]
  \vspace*{-0.3cm}
  \begin{minipage}[t]{7.0cm}
    \vspace*{-6.4cm}
    \epsfig{file=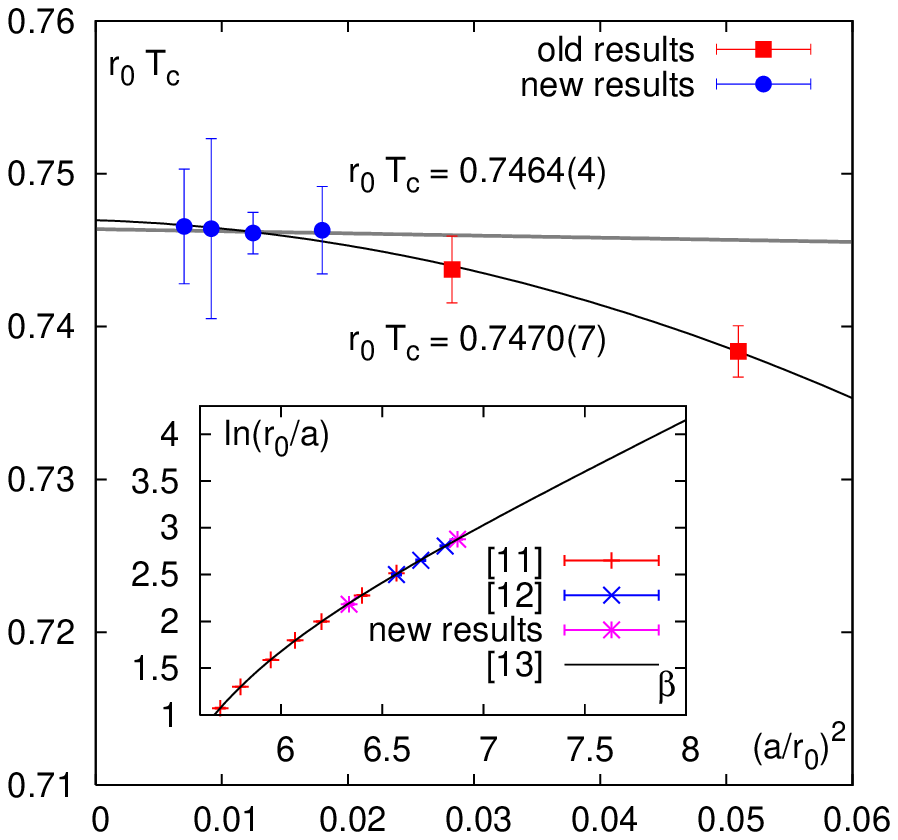,width=7.0cm}
  \end{minipage}\hspace*{7mm}%
  \begin{minipage}[t]{7.0cm}
    \hspace*{-1.6cm}
    \epsfig{file=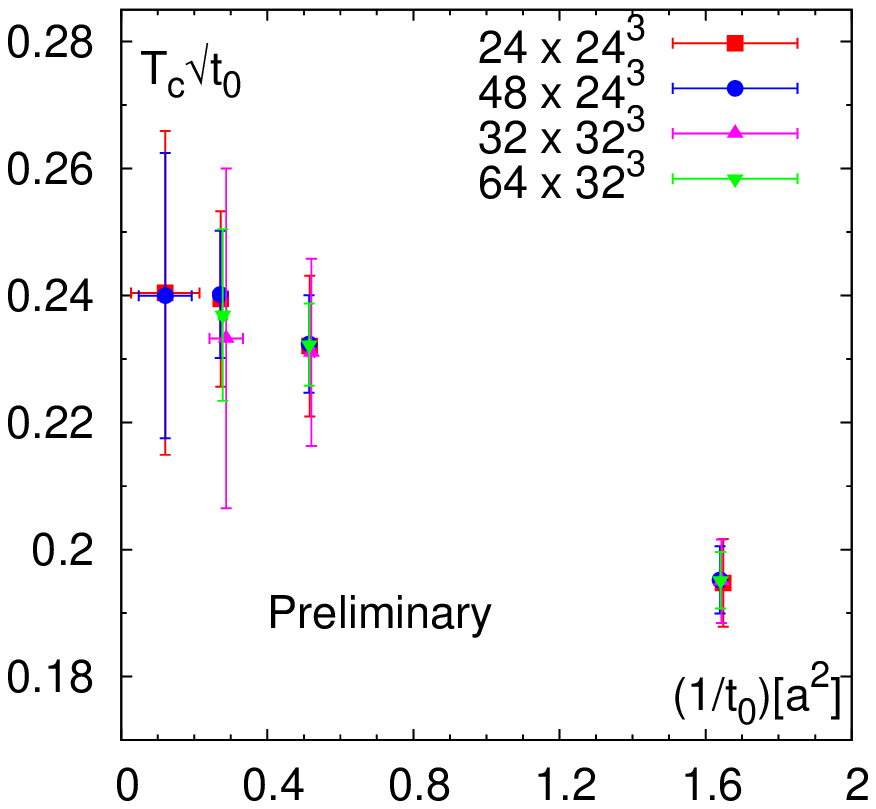,width=9.3cm}
  \end{minipage}
\caption[a]{
Left: Continuum extrapolation for 
 $r_0 \Tc$. The conversion
 from $\beta_c$ to $r_0 / a$ is based on 
 refs.~\cite{pos18,pos19} and additional 
 new simulations, together with a rational fit from 
 ref.~\cite{pos20} (inset).
 The result $r_0 \Tc = 0.7470(7)$
 can be contrasted with $r_0 \Tc = 0.7498(50)$ from ref.~\cite{sn}.
 (For comparisons with perturbation theory, 
 $r_0 \Lambdamsbar = 0.602(48)$ from ref.~\cite{pos22} yields
 $\Tc/\Lambdamsbar = 1.24(10)$; 
 ref.~\cite{pos23} suggests $r_0 \Lambdamsbar \simeq 0.637(32)$
 which would yield $\Tc/\Lambdamsbar \simeq 1.17(6)$.)
 Right: 
 Preliminary results for $\sqrt{t_0}T_c $ as a function of $a^2/t_0$. 
}
\label{fig:Tc}
\end{figure}
%

Considering first the Sommer scale, 
the data of \fig\ref{fig:betac}(right) can be used for determining 
the dimensionless combination $r_0 \Tc$; 
the results are shown in \fig\ref{fig:Tc}(left).
The results display the expected {\cal O}($a^2$) behaviour
and can be extrapolated to the continuum. We obtain $r_0\Tc \simeq 0.7470(7)$
which agrees with an earlier result from ref.~\cite{sn} 
but displays a much reduced error. 

We also have preliminary estimates for $\sqrt{t_0} \Tc$, 
obtained at different lattice extensions from zero-temperature simulations
at $\beta=\beta_c(N_\tau=4,6,8,12)$,  
and shown in \fig\ref{fig:Tc}(right). 
Increasing the statistics and number of lattice ensembles 
hopefully yields accurate results for $\sqrt{t_0}T_c$.

%
\section{Application to heavy quark diffusion} %
\la{se:GE}

\begin{figure}[t]

  \begin{minipage}[t]{8.5cm}
    \vspace*{-1cm}
    \epsfig{file=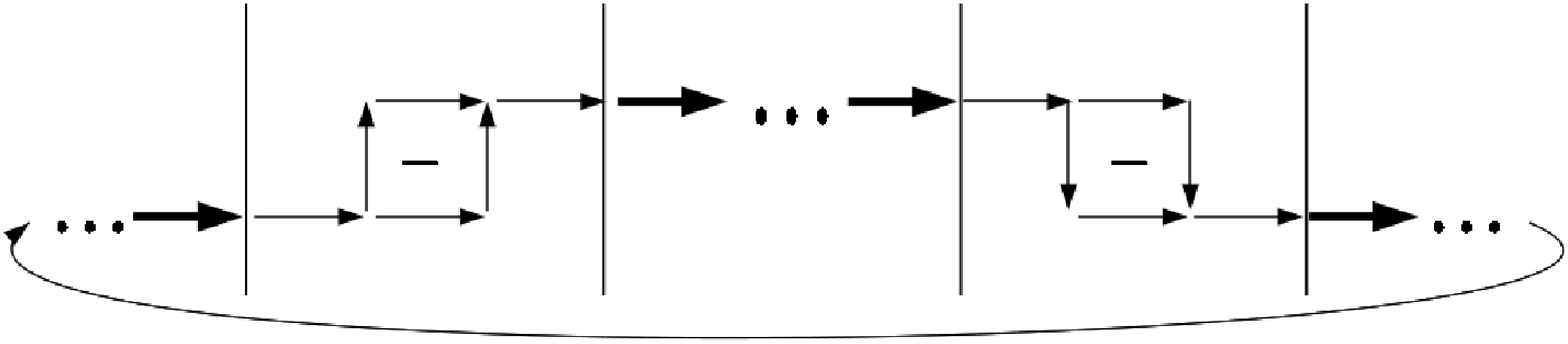,angle=0,width=7.5cm}
  \end{minipage}
  \begin{minipage}[t]{7.0cm}
    \small
    \begin{tabular}{|c|c|c|c|c|c|} 
    \hline      
       $\beta$ & $N_\tau$ & $N_s$ & $N_{\mathrm{conf}}$ & $N_{\mathrm{stat}}$
       & $r_0 T$  \\
    \hline\hline
     6.872 & 16 & ~32 & 139 &  1000 &   1.111 \\
     6.872 & 16 & ~64 & 99  &  1000 &   1.111 \\
     7.192 & 24 & ~96 & 159 &  1000 &   1.077 \\
     7.544 & 36 & 144 & 278 &  1000 &   1.068 \\
    \hline                         
    \end{tabular}
  \end{minipage}

\vspace*{3mm}

\caption[a]{
Left: 
 Fat links, thin links and electric fields along the time direction
 (cf.\ the text).
Right:
 Run parameters. 
 The values of $r_0 T$ are obtained through an interpolation/extrapolation
 as illustrated in \fig\ref{fig:Tc}(left). With the value of 
 $r_0 \Tc$ from \fig\ref{fig:Tc}(left), we have $T / \Tc \approx 1.43$
 for $N_\tau = 36$.
}
\label{fig:multilevel-linkintegrated}
\label{tab:statistics}
\end{figure}

Heavy quarks carry a colour charge and, whenever there are gauge 
fields present, are subject to a coloured Lorentz force, which 
adjusts their velocities to those corresponding to kinetic equilibrium
(this corresponds to the physics of diffusion). 
Through linear response theory the effectiveness of the adjustment can 
be related to a ``colour-electric correlator''~\cite{pos4,eucl},  
\be
 G_\rmi{\,E}(\tau) \equiv - \fr13 \sum_{i=1}^3 
 \frac{
  \Bigl\langle
   \re\tr \Bigl[
      U(\fr{1}{T};\tau) \, gE_i(\tau,\vec{0}) \, U(\tau;0) \, gE_i(0,\vec{0})
   \Bigr] 
  \Bigr\rangle
 }{
 \Bigl\langle
   \re\tr [U(\fr{1}{T};0)] 
 \Bigr\rangle
 }
 \;, \la{GE_final}
\ee
where $gE_i$ denotes the  
colour-electric field, $T$ the temperature, and $U(\tau_2;\tau_1)$
a Wilson line in the Euclidean time direction. 
A discretized version of this correlator 
is shown in \fig\ref{fig:multilevel-linkintegrated}.

Preliminary lattice measurements of the colour-electric correlator
have already been carried out~\cite{hbm,mumbai,measure}. 
The results look promising, 
hinting at a large and phenomenologically interesting 
non-perturbative effect in the diffusion coefficient. 
However, none of these results 
contain a systematic continuum extrapolation. The ultimate goal of 
our investigation is to perform one, by making use of the results of
\ses\ref{se:betac}, \ref{se:Tc} as well as of new simulations.  

\begin{figure}[t]
  \vspace*{-0.5cm}
  \begin{minipage}[t]{7.0cm}
    \vspace*{0.0cm}
    \epsfig{file=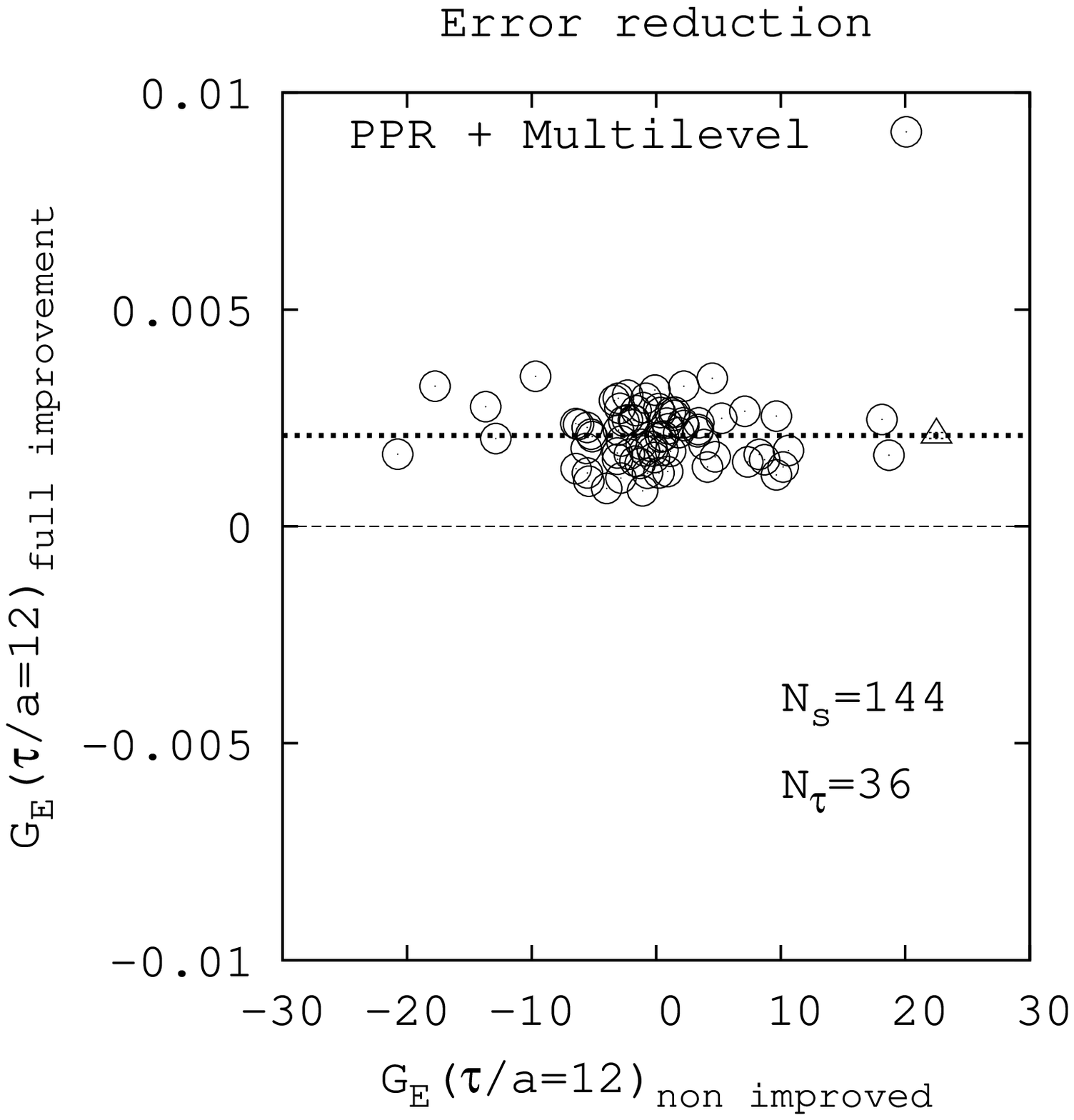,angle=270,width=6.4cm}
  \end{minipage}
  \begin{minipage}[t]{9.25cm}
    \vspace*{0.55cm}
    \epsfig{file=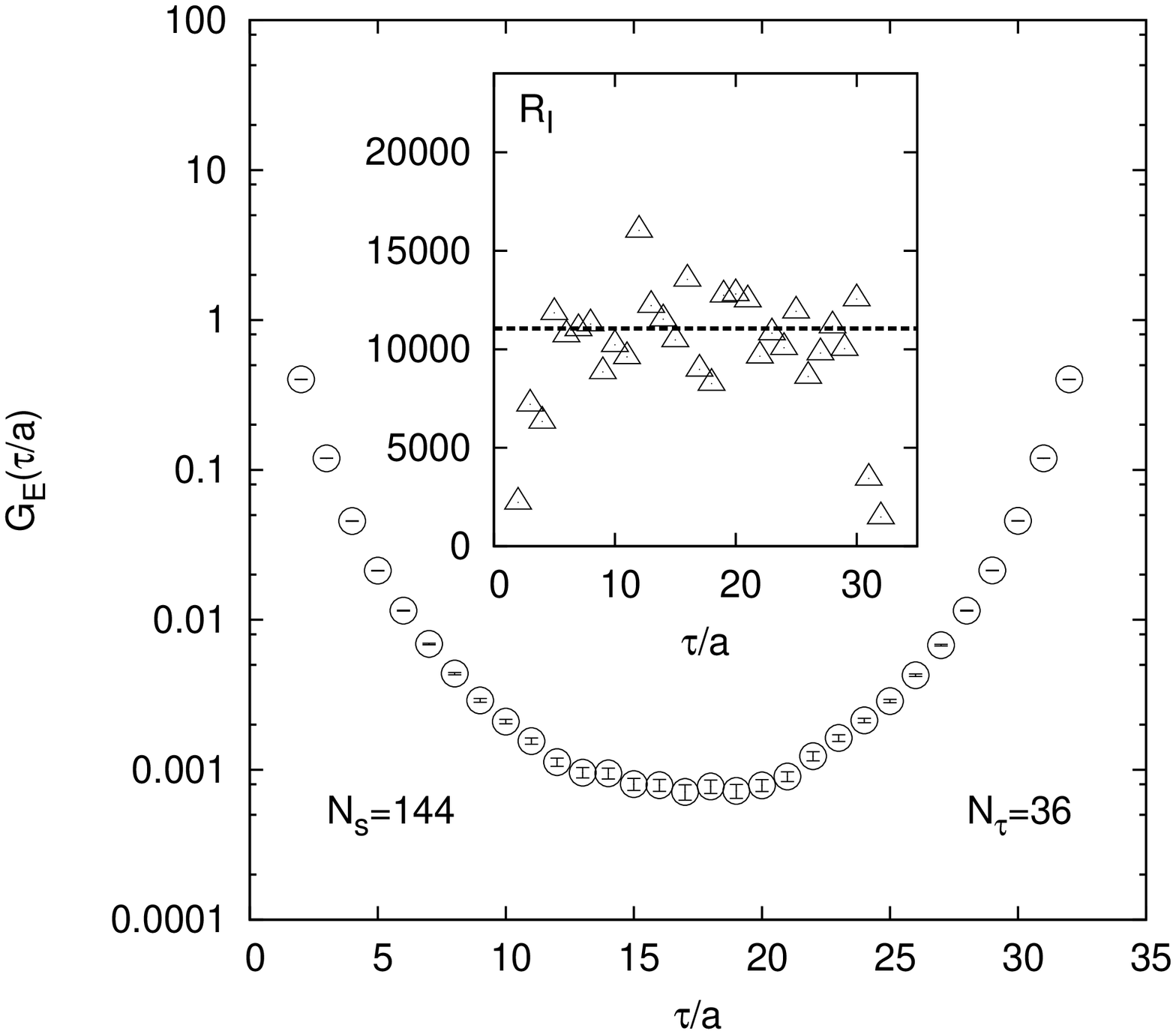,angle=270,width=8.5cm}
  \end{minipage}
\caption[a]{
Left: 
 Error reduction for $G_\rmii{\,E}(\tau=12 a)$ 
 in a $36 \times 144^3$ box with techniques described 
 in the text. 
 For $77$ statistically independent 
 configurations we determine the non-improved observable ($x$-axis) and
 the fully improved one ($y$-axis). A dotted horizontal line and 
 a triangle mark 
 the average of the improved observable.
Right: 
 The statistically improved correlator $G_\rmii{\,E}(\tau)$. 
 The inset shows the ratio of non-improved over improved statistical errors, 
 denoted by $R_{\rm I}$. An error reduction by a factor 
 $R_{\rm I}^{ } \sim {\cal O}(10^{4})$ can be achieved.
}
\label{fig:error}
\end{figure}

Our numerical investigations are based on standard lattice QCD
techniques. We employ the Wilson gauge action on fine 
isotropic lattices with lattices spacings down to $\sim 0.015$~fm.
The measurement of \eq\nr{GE_final}
is performed on gauge field
configurations generated using 500 sweeps between subsequent
configurations, guaranteeing that the configurations are statistically
independent. However 
the correlation function in \eq\nr{GE_final} decreases rapidly with
$\tau$, and for $\tau \sim \frac{1}{2T}$ suffers from a 
weak signal-to-noise ratio. In order to tackle this problem we  
use a multi-level update \cite{lw,shear} for the part of the 
operator that includes the electric field insertions, and 
link-integration (``PPR'') \cite{ppr,fr} for the straight lines between
them (the ``fat links'' in \fig\ref{fig:multilevel-linkintegrated}). 
The table in \fig\ref{tab:statistics} summarizes 
the situation with $N_{\mathrm{conf}}$
labelling the number of statistically independent configurations 
and $N_{\mathrm{stat}}$ the number of additional 
``multilevel'' updates.
As demonstrated in \fig\ref{fig:error}, these techniques 
suffice to yield a signal.

%
%

The lattice sizes of our simulations, 
corresponding to a temperature $T \sim 1.43 T_c$, 
substantially exceed those of 
earlier simulations~\cite{hbm,mumbai,measure}, 
which had $V_{\rm max}=24 \times 64^3$.
Due to the increased number of multilevel updates 
we focussed on lattices with an aspect
ratio $N_s/N_\tau=4$ (cf.\ the table in \fig\ref{tab:statistics}), 
which was the optimal choice given 
the available computational resources.
However, volume-scaling checks have been 
performed at the smallest values of $N_\tau$.

%
%

\begin{figure}[t]

\begin{center}
  \epsfig{file=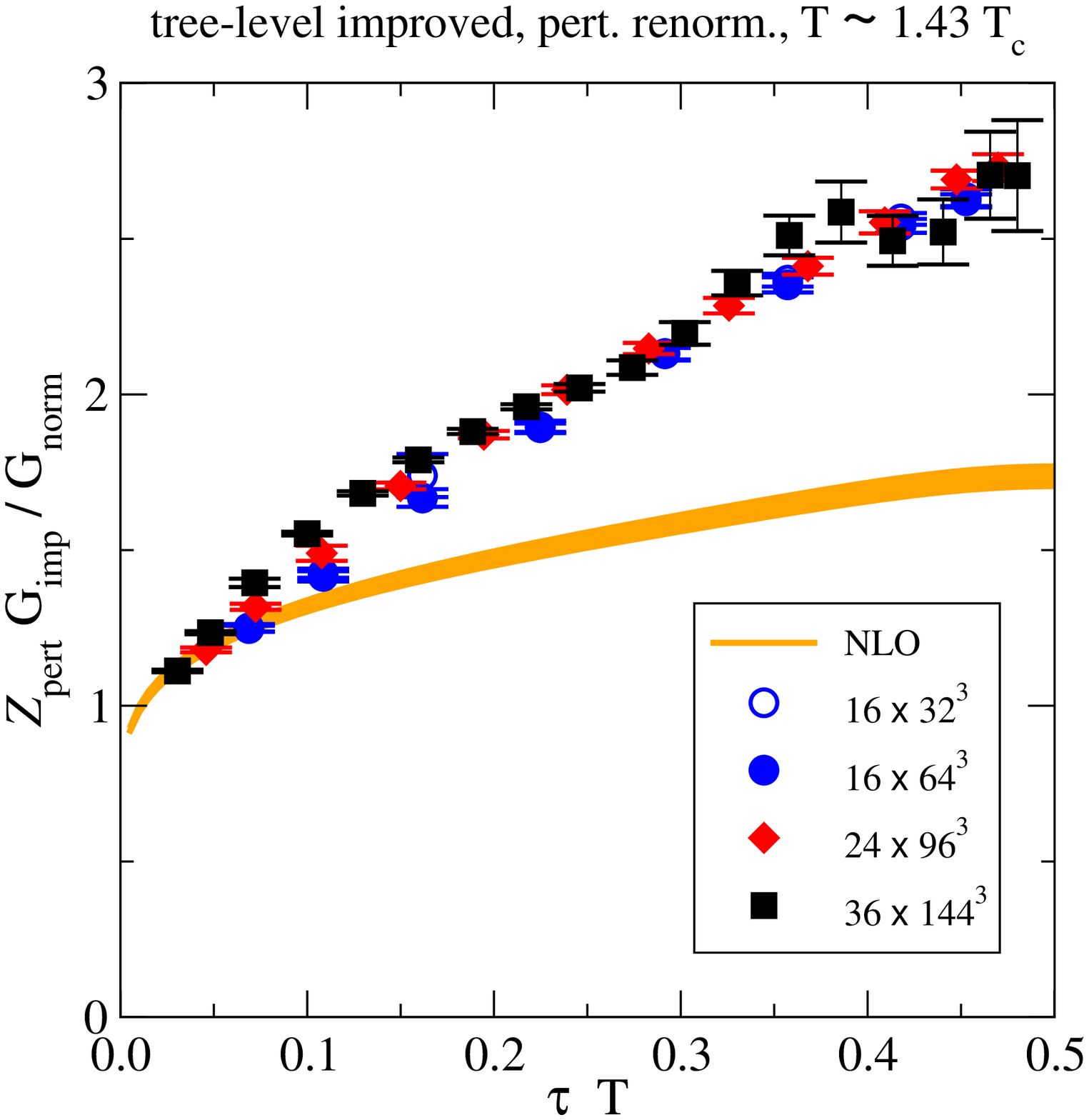,width=7.0cm}
\end{center}

\vspace*{-3mm}

\caption[a]{
$G^{ }_\rmii{E}$ 
at $T\sim 1.43 T_c$. For the NLO result we have varied
$\Tc/\Lambdamsbar\in(1.11,1.34)$, cf.\ 
\fig\ref{fig:Tc}.}
\label{fig:imp_norm_wZ_pert}
\end{figure}

After tree-level improvement~\cite{rs,shear} our measurements  
yield a correlator denoted by $G_\rmi{\,imp}(\tau)$, which 
is furthermore multiplied by a perturbative renormalization
factor ${Z}_\rmi{pert}$~\cite{measure}. Normalizing the resulting 
correlator to  
\be
 G_\rmi{norm}(\tau T)
 \; \equiv \; 
 \pi^2 T^4 \left[
 \frac{\cos^2(\pi \tau T)}{\sin^4(\pi \tau T)}
 +\frac{1}{3\sin^2(\pi \tau T)} \right] 
 \;,
\ee
the data are displayed in \fig\ref{fig:imp_norm_wZ_pert}. They 
exhibit a clear enhancement over 
the next-to-leading order (NLO) prediction from ref.~\cite{rhoE}.
A continuum extrapolation remains to be carried out.

%
\section{Outlook}

The Euclidean correlator of \eq\nr{GE_final}
is related to a corresponding spectral function
$ \rho_\rmii{\,E}$ through
$ 
 G_\rmii{\,E}(\tau) =
 \int_0^\infty
 \frac{{\rm d}\omega}{\pi} \rho_\rmii{\,E}(\omega)
 \frac{\cosh \left(\frac{1}{2} - \tau T \right)\frac{\omega}{T} }
 {\sinh\frac{\omega}{2 T}} 
$.
Once a perturbatively determined short-distance divergence
is subtracted from a continuum-extrapolated $G_\rmii{\,E}(\tau)$, 
the remainder may be subjected to 
an analytic continuation algorithm~\cite{cuniberti,analytic} 
or a well-motivated 
model like in refs.~\cite{mumbai,measure}. 
In particular, 
a ``momentum diffusion coefficient'', often denoted by $\kappa$, can be 
obtained from 
$
 \kappa = \lim_{\omega\to 0} \frac{2 T \rho_\rmii{\,E}(\omega)}{\omega}
$. 
In the non-relativistic limit 
(i.e.\ for $M \gg \pi T$, where $M$ stands for a heavy quark mass)
the corresponding ``diffusion coefficient'' 
is given by $D = 2 T^2/\kappa$. 
It will be interesting to see 
whether preliminary estimates of the diffusion 
coefficient~\cite{mumbai,measure}
can be confirmed after a continuum limit, and how well the results perform in
phenomenological comparisons with LHC heavy ion data 
(cf.\ e.g.~ref.~\cite{pos26}). 
Of course, taking the continuum limit plays an important role in the extraction
of the light-quark diffusion coefficient
and electrical conductivity as well~\cite{ding1,cond}.

\section*{Acknowledgment}                                    
                       
This work has been supported
in part by the DFG under grant GRK881, 
by the SNF under grant 200021-140234, 
and by the European Union through HadronPhysics3 and ITN STRONGnet
(grant 238353).
Simulations were
performed using JARA-HPC resources at the RWTH Aachen
(project JARA0039), JUDGE/JUROPA at the JSC J\"ulich, 
the OCuLUS Cluster at the Paderborn Center for Parallel
Computing, and the Bielefeld GPU cluster.

%
%

%

\end{document}